\documentclass[aps,twocolumn,pra,superscriptaddress,showpacs,tightenlines]{revtex4}
\usepackage{mathrsfs}
\usepackage{amssymb}
\usepackage{amsmath}
\usepackage{graphicx}
\usepackage{epsfig}
\usepackage{txfonts}
\usepackage{subfigure}
\usepackage{amsfonts}

\begin{document}
\title{Complementarity via error-free measurement in a two-path interferometer}

\author{Yanjun Liu}
\affiliation{Key Laboratory of Low-Dimensional Quantum Structures and Quantum Control of Ministry of Education,
Department of Physics and Synergetic Innovation Center of Quantum Effects
and Applications, Hunan Normal University, Changsha 410081, China}
\author{Jing Lu}
\affiliation{Key Laboratory of
Low-Dimensional Quantum Structures and Quantum Control of Ministry of Education,
Department of Physics and Synergetic Innovation Center of Quantum Effects
and Applications, Hunan Normal University, Changsha 410081, China}
\author{Lan Zhou}
\thanks{Corresponding author}
\email{zhoulan@hunnu.edu.cn} \affiliation{Key Laboratory of Low-Dimensional Quantum Structures and Quantum Control of Ministry of Education,
Department of Physics and Synergetic Innovation Center of Quantum Effects
and Applications, Hunan Normal University, Changsha 410081, China}

\begin{abstract}
We study both the wave-like behavior and particle-like behavior in a general Mach-Zehnder interferometer with its asymmetric beam splitter.
A error-free measurement in the detector is used to extract the which-path information. The fringe visibility V
and the which-path information $I_{path}$ are derived: their complementary relation $V+I_{path}\leq1$ are found, and the condition for the equality is also presented.
\end{abstract}

\pacs{03.67.-a, 03.65.Ta, 07.60.Ly}

\maketitle \narrowtext


\section{Introduction}
The wave-like nature and particle-like nature are two mutual exclusion properties of quantum systems, and the appearance of these two properties 
are determined by the experimental instrument, which is known as Bohr's complementarity principle \cite{Bohr}. The wave-particle duality is the well-known 
example used to exhibit the curious nature of the complementarity principle. In the recoiling-slit gedanken experiment introduced by Einstein and Bohr where a particle is sent through a movable 
slit placed before a double slit, Wootters and Zurek \cite{Woo} proposed their quantitative formulation of the wave-particle duality. In a two-path interferometer, such as Mach-Zehnder interferometer (MZI) \cite{Wolf},
a complementarity was first found between a priori fringe visibility of the interference pattern and the predictability \cite{Yasin}, which were determined by the initial state of the particle.
Later on, the wave-like and particle-like nature were characterized by the visibility of the interferometer fringe and the path distinguishability \cite{Englert}, respectively, and there is a trade-off between these quantities. Both predictability and distinguishing are quantities that measure the which-path knowledge. Since there are many ways to define the measure of which-path knowledge, the complementarity 
between the fringe visibility and the which-path knowledge has been studied greatly in theory and experiment \cite{Yasin,Mandel,Jaeger,Horne,Englert,Rempe,rrS,Zawisky,Kaszlikowski,Bimonte,Bramon,Jakob,BanO,Masashi,
Han,LiLi,Fonseca,Asad,Angelo}.

Form the information theory viewpoint, the achievement of knowledge is an information transmission process. It is natural to use information measure to characterize the particle
nature of a quantum system in a two-path interferometer. Here, we employ the mutual information which is called which-path information (WPI). The information cannot be transmitted 
until a measurement is performed, for example, a detector is placed in one path of the MZI in Ref.\cite{Englert}, and an error-minimum state distinguishing measurement \cite{Helstrom}
is performed on the detector after the particle interacts with the detector to acquire the path distinguishability.
Although such ambiguous measurement yields a conclusive outcome, there is nonvanishing probability of making a wrong guess since errors in the conclusive outcome are unavoidable. The 
other optimized measure strategies is the error-free discrimination \cite{Chefles} among nonorthogonal states, which allow a nonzero probability of inconclusive outcomes.

In this paper, we study the trade-off between the fringe visibility and WPI in a two-path interferometer made of one symmetric beam splitter (BS) and one asymmetric BS. The 
error-free measurement is used to obtain the WPI. It is found that the magnitudes of the fringe visibility and the WPI are effected by the asymmetric BS and the input state
of the particle. A bound between the fringe visibility and the WPI is also found.    

The paper is organized as follows. In Sec. \uppercase\expandafter{\romannumeral2}, we introduce the setup that a quantum system display its wave-like behavior and particle-like behavior.
In Sec. \uppercase\expandafter{\romannumeral3}, the WPI is defined, and an unambiguous discrimination on the state of the detector is used to obtain the which-path information.
In Sec. \uppercase\expandafter{\romannumeral4}, we made our conclusion.


\section{\label{Sec:2}The setups and the state evolution}

\begin{figure}[tbp]
\includegraphics[clip=true,height=5cm,width=8cm]{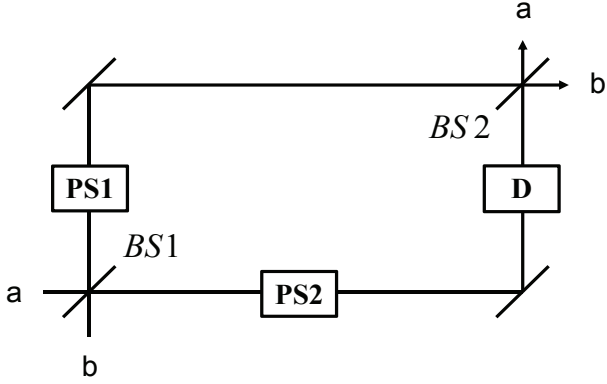}
\caption{The schematic sketch of the general Mach-Zehnder interferometer with the second BS asymmetric, a detector is placed in path a.} \label{fig1a.eps}
\end{figure}

A general MZI, shown in Fig.~\ref{fig1a.eps}, consists of two BSs and phase shifters (PSs). A beam of particles coming from either port a or b is
first splitted by beam splitter BS1 into two beams and then these beams are recombined by BS2. So two paths a and b are available between BS1 and BS2. A particle taking
path a(b) is denoted by the state $|a\rangle=\hat{a}^{\dagger}|0\rangle$ $(|b\rangle=\hat{b}^{\dagger}|0\rangle)$, where $\hat{a}^{\dagger}$ and $\hat{b}^{\dagger}$
are the corresponding creation operators in path a and b, and they satisfy $[\hat{a},\hat{b}]=0$,  $[\hat{a},\hat{a}^{\dagger}]=1$,  $[\hat{b},\hat{b}^{\dagger}]=1$. States $|a\rangle$ and $|b\rangle$
support a two-dimensional $H_{q}$. In this sense, a quantum bit is formed. The state of particle traveling in this interferometer is characterized by the change of a Bloch vector in a Bloch sphere. Before the particle
incident into the general MZI, the state of the particle is described by the density matrix
\begin{equation}
\rho _{in}^{Q}=\frac{1}{2}( 1+S_{x}\sigma _{x}+S_{y}\sigma
_{y}+S_{z}\sigma _{z}),
\end{equation}%
where $\sigma _{x},\sigma _{y},\sigma _{z}$ are the Pauli matrix and $\sigma_{z}=|b\rangle\langle b|-|a\rangle\langle a|$. Here, the initial Bloch vector $\overrightarrow{S}=(\overrightarrow{S_{x}},\overrightarrow{S_{y}},\overrightarrow{S_{z}})$.
When $|\overrightarrow{S}|=1$, the particle is in a pure state, when $|\overrightarrow{S}|<1$ , the particle is in a mixed state.

The effect of the BS on the state of the incoming particles is described by the operator
$B(\beta)=\exp[i\beta(\hat{a}^{\dagger}\hat{b}+\hat{b}^{\dagger}\hat{a})]$, which preserves the total number of particles in this MZI. In the subspace spanned by basis $\{|a\rangle,|b\rangle\}$, the BS performs a rotation
around the y axis by angle $\beta$, which is denoted by
\begin{equation}
U_{B}(\beta)=\exp ( -i\frac{\beta }{2}\sigma _{y})
=(
\begin{array}{cc}
\sqrt{t} & -\sqrt{r} \\
\sqrt{r} & \sqrt{t}
\end{array}).
\end{equation}
Here, $r$ and $t$ respectively represent the reflection coefficient and transmission coefficient of the beam splitter.
\begin{equation}
r=\sin^{2}{\frac{\beta}{2}} ,\text{ \ \ \ \ \ \ \ \ }  t=\cos^{2}{\frac{\beta}{2}}.
\end{equation}

The PS in path $d \in \{a,b \}$ is described by $P(\phi_{d})=\exp(i\phi_{d}\hat{d}^{\dagger}\hat{d})$.
If the parameters $\phi_{a}$ and $\phi_{b}$ have the same magnitude but different sign, i.e., $\phi_{a}=-\phi_{b}=\phi$, a rotation around the $z$ axis
by angle $\phi$ is realized by PS1 and PS2,
\begin{equation}
U_{P}(\phi)=\exp ( -i\frac{\phi}{2}\sigma _{z}).
\end{equation}

To acquire the WPI, a detector is usually placed in one of the paths (e.g. the a path). As long as the particle go through the general MIZ, a operator
\begin{equation}
M=\frac{1+\sigma_{Z}}{2}I+\frac{1+\sigma_{Z}}{2}U,
\end{equation}
performs on the detector, where U is unitary.
The final state of the particle and the detector reads
\begin{eqnarray}
\rho _{f}&=&U_{B}(\beta)MU_{P}(\varphi)U_{B}(\frac{\pi}{2})\rho_{in}^{Q}\rho_{in}^{D}
U_{B}^{\dagger}(\frac{\pi}{2})U_{P}^{\dagger}(\varphi)M^{\dagger}U_{B}^{\dagger}(\beta) \notag\\
&=&\frac{1}{4}( 1-S_{x}) ( 1+\sigma _{z}\cos \beta
+\sigma _{x}\sin \beta ) \otimes \rho _{in}^{D} \notag\\
&&-\frac{1}{4}e^{-i\phi
}( S_{z}-iS_{y}) ( \sigma _{z}\sin
\beta -\sigma _{x}\cos \beta -i\sigma _{y}) \otimes \rho _{in}^{D}U^{\dagger} \notag\\
&&-\frac{1}{4}e^{i\phi}( S_{z}+iS_{y}) ( \sigma _{z}\sin \beta-\sigma _{x}\cos
\beta  +i\sigma _{y}) \otimes U\rho _{in}^{D}\notag\\
&&+\frac{1}{4}( 1+S_{x}) ( 1-\sigma _{z}\cos
\beta -\sigma _{x}\sin \beta ) \otimes
U\rho _{in}^{D}U^{\dagger},
\end{eqnarray}
where $\rho_{in}^{D}$ is the initial state of the detector, and the BS1 is assumed to be symmetric.

The probability that we finding the particle at the output port $a$ reads
\begin{eqnarray}
p(\phi)&=&tr_{QD}[ \frac{1}{2}( 1-\sigma _{z}) \rho _{f}]  \notag\\
&=& \frac{1}{2}( 1+ S_{x}\cos\beta)  \notag\\
&&+\frac{1}{2}\sqrt{S_{z}^{2}+S_{y}^{2}}\sin\beta
|tr_{D}(U\rho _{in}^{D})| \cos ( \alpha +\gamma +\phi )  \label{2eq-01},
\end{eqnarray}
where $\alpha$ and $\gamma$ are defined as
\begin{equation}
\alpha= \arctan\frac{S_{y}}{S_{z}},  \text{ \ \ \ \ \ \ }
\gamma= -i\ln\frac{tr_{D}(U\rho_{in}^{D})}{|tr_{D}(U\rho_{in}^{D})|}.
\end{equation}
The fringe visibility which documents the wave-like property of the particle is defined via the probability in Eq.~(\ref{2eq-01}) as
\begin{eqnarray}
V&=&\frac{max P(\phi)-min P(\phi)}{max P(\phi)+min P(\phi)} ,
\end{eqnarray}
where the maximum and minimum is achieved by adjusting $\phi$. And one can easily obtain that
\begin{eqnarray}
V&=&\frac{\sin
\beta }{1+S_{x} \cos \beta }\sqrt{S_{z}^{2}+S_{y}^{2}}| tr_{D}(
U\rho_{in} ^{D})| \label{2eq-02}.
\end{eqnarray}
We note that the fringe visibility can also be defined by the probability at the output port $b$. However, the fringe visibility measured
in either output port $a$ or $b$ is different expect the BS2 is symmetric (i.e. $\beta=\pi/2$).

\begin{figure*}[htbp]
\includegraphics[clip=true,height=8cm,width=14cm]{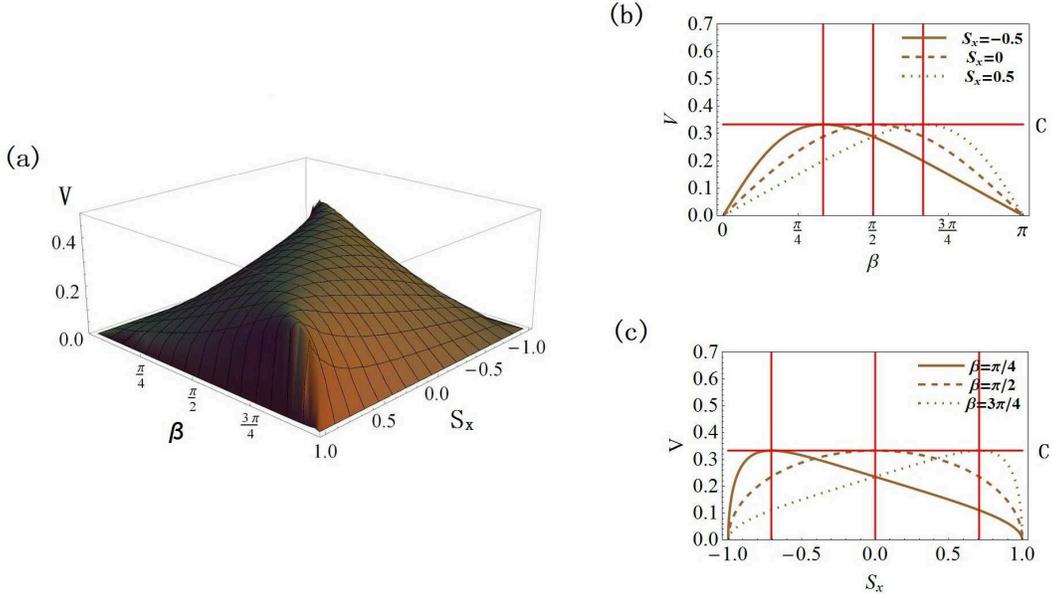}
\caption{(Color online).(a) The fringe visibility as the function of the $S_{x}$ and $\beta$ with $| tr_{D}(U\rho_{in} ^{D})|=1/3$ and $S_{x}^{2}+S_{y}^{2}+S_{z}^{2}=1$,                                                                                                                                  (b) the cross section of (a) when $S_{x}=-0.5, 0, 0.5$, (c) the cross section of (a) when $\beta=\Pi/4, \Pi/2, 3\Pi/4$. } \label{fig2.eps}
\end{figure*}

Eq.~(\ref{2eq-02}) shows that both the BS2 and the initial state have influence on the fringe visibility. It can be found that for a given $\beta$, more wave nature
appears when the particle is in a pure state $(|\overrightarrow{S}|=1)$. To show the dependence of the wave nature of the BS2 and the initial state, we have plotted the
fringe visibility V as a function of the $\beta$ and $S_{x}$ for $| tr_{D}(U\rho_{in} ^{D})|=1/3$ and $|\overrightarrow{S}|=1$ in Fig.~\ref{fig2.eps}. It can be observed for a given $S_{x}(\beta)$,  the
fringe visibility first increase and then decrease as $\beta(S_{x})$ increases, and the fringe visibility obtains the maximum
$C \equiv |tr_{D}(U\rho_{in} ^{D})|$ when $\cos \beta=-S_{x}$. This is the reason the quantified wave-particle duality in \cite{Englert} is presented by choosing $S_{x}=0$ when $\beta=\pi/2$.
The value of the fringe visibility is zero in the following situation: (1) The effect of the BS2 for the particle is full transmission or full reflection, corresponding to $\beta=0$ or $\pi$. (2)
The particle travels only in a path or b path, corresponding to $S_{x}=1$ or $-1$.

\begin{figure}[tbp]
\includegraphics[clip=true,height=5cm,width=8cm]{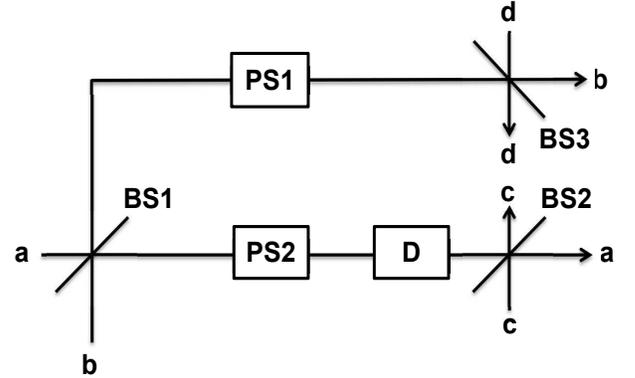}
\caption{Schematic representation of the setup with four input and output ports.} \label{fig1b.eps}
\end{figure}

It is well-known that the wave-like property characterized by the fringe visibility is complementary to the particle-like property which
gives rise to the WPI. A decrease of the fringe visibility predicts an increasing of the WPI.
From Eq.~(\ref{2eq-02}), one can find that the BS2 besides the initial state affect
the WPI, which indicates that the asymmetric BS2 introduce additional WPI \cite{LiLi}. Then, the set up
of measuring the particle-like property is different form the one which is obtained simply by removing the BS2 in Fig.~\ref{fig1a.eps}. Actually, the particle-like property is measured by the setup with four input and output
ports, which is shown in Fig.~\ref{fig1b.eps}. Since two paths c and d have been introduced, the initial density matrix for the total
system reads
\begin{gather}
\rho _{in}^{QD}=\rho _{in}^{Q}\otimes|00\rangle_{cd}\langle00|\otimes\rho _{in}^{D},
\end{gather}
where $|00\rangle_{cd}$ is vacuum states of the input ports c and d. Before the particle meets
the BS2 and BS3, the state of the particle and the detector is the same to the state before the BS2 in Fig.~\ref{fig1a.eps}.
The BS2 acts on the paths a and c, and BS3 acts on the paths b and d. The performance of BS2 and BS3 is denoted by
$B_{2}=exp[(-\frac{\beta}{2})(\hat{a}^{\dagger}\hat{c}-\hat{c}^{\dagger}\hat{a})]$ and
$B_{3}=exp[(-\frac{\beta}{2})(\hat{b}^{\dagger}\hat{d}-\hat{d}^{\dagger}\hat{b})]$ respectively. After the particle goes through the BS2 and BS3, the
state for the particle appearing in either output a or d reads
\begin{eqnarray}
\rho^{QD} _{f}&=&\omega_{b} |d\rangle \langle d|
\rho _{in}^{D}
+\omega_{a} | a\rangle
\langle a|  U\rho _{in}^{D}U^{\dagger} \notag\\
&&+\frac{\sqrt{rt}}{1+S_{x}(t-r)}e^{i\phi}( S_{z}+iS_{y} ) | a\rangle
\langle d| U\rho _{in}^{D}  \notag\\
&&+\frac{\sqrt{rt}}{1+S_{x}(t-r)}e^{-i\phi}(S_{z}-iS_{y})| d\rangle \langle a|
\rho _{in}^{D}U^{\dagger}  \label{2eq-04}.
\end{eqnarray}
Here, the \textit{prior} probabilities
\begin{equation}
\omega _{a} =\frac{t( 1+S_{x}) }{1+S_{x}(t-r) }, \text{ \ \ \ \ \ \ } \omega _{b} =\frac{r( 1-S_{x}) }{1+S_{x}(t-r)},
\end{equation}
for finding the particle in output a and d respectively. Since the particle detected at output port a(d) is
obtained by transmission (reflection) from path a(b), we change the letter d in Eq.~(\ref{2eq-04}) by b, which is rewritten as
\begin{eqnarray}
\rho^{QD} _{f}&=&\omega_{b} |b\rangle \langle b|
\rho _{in}^{D}
+\omega_{a} | a\rangle
\langle a|  U\rho _{in}^{D}U^{\dagger} \notag\\
&&+\frac{\sqrt{rt}}{1+S_{x}(t-r)}e^{i\phi}( S_{z}+iS_{y} ) | a\rangle
\langle b| U\rho _{in}^{D}  \notag\\
&&+\frac{\sqrt{rt}}{1+S_{x}(t-r)}e^{-i\phi}(S_{z}-iS_{y})| b\rangle \langle a|
\rho _{in}^{D}U^{\dagger}  \label{2eq-05}.
\end{eqnarray}


\section{\label{Sec:3} Information gain via error-free measurement }


After tracing over the degree of the particle in Eq.~(\ref{2eq-05}), we obtain the final state of the detector
\begin{gather}
\rho^{D} _{f}= \omega_{b}\rho _{in}^{D}
+\omega_{a}\rho _{out}^{D}.
\end{gather}

To obtain the WPI, we have to discriminate the states $\rho _{in}^{D}$ and $\rho _{out}^{D}\equiv U\rho _{in}^{D}U^{\dagger}$ with prior probabilities $\omega _{b}$ and $\omega _{a}$
in an optimal way. Here, we perform the error-free measurement on the detector. This kind of measurement gives two results: a conclusive one without any error and an inconclusive
one. The conclusive result means which-path the particle takes is definitely known. Mathematically, to calculate the WPI,
one has to introduce the positive operator-valued measure (POVM) $\{{\Pi _{k},k=a,b,0}\}$ with the resolution of the identity $\sum_{k}\Pi _{k}=I$, which leads to an inconclusive outcome 0 and two definitely results
a and b. The unambiguous discrimination requires
\begin{equation}
\rho _{in}^{D}\Pi_{a}=\rho_{out} ^{D}\Pi_{b}=0  \label{3eq-01}.
\end{equation}
The joint probability that the particle travels on path $d \in \{a,b \}$ and the which-path result k indicated from the measurement of the detector reads
\begin{eqnarray}
Q(\mu,k)&=&Tr_{D}\langle \mu|\Pi_{k}\rho_{f}^{QD}|\mu\rangle  \label{3eq-02}.
\end{eqnarray}
Then, the amount of the WPI \cite{Thomas} obtained from the error-free measure is given by
\begin{eqnarray}
I_{path} &=&\underset{\mu =a,b}{\sum }\underset{k=a,b}{\sum }Q( \mu
,k) \log [ \frac{Q( \mu ,k) }{Q( \mu )
Q( k) }]     \label{3eq-03}.
\end{eqnarray}

For the sake of simplicity, we assume that the detector is initially in a pure state $\rho _{in}^{D}=| r\rangle\langle r|$. Since the unitary operator U is arbitrary, states $| r\rangle$ and $| s\rangle \equiv U| r\rangle$
can be assumed to be linearly independent, the POVM is constructed as
\begin{eqnarray}
\Pi _{a} &=&\alpha | r^{\perp}\rangle \langle r^{\perp}| ,  \notag\\
\Pi _{b} &=&\beta | s^{\perp}\rangle\langle s^{\perp}| ,  \notag \\
\Pi _{0}&=&( 1-\beta S^{2}) |r \rangle \langle r | +\beta SC| r\rangle \langle  r^{\perp}| \notag\\
&&+\beta SC|  r^{\perp}\rangle \langle
r| +( 1-\beta C^{2}-\alpha) |
\ r^{\perp}\rangle \langle  r^{\perp}| ,\label{3eq-04}
\end{eqnarray}
where states $| r\rangle(| s\rangle)$ are orthogonal to
\begin{eqnarray}
| r^{\perp}\rangle  &=&\frac{1}{S}( |s\rangle -C| r\rangle ),   \notag\\
| s^{\perp}\rangle  &=&\frac{1}{S}( |r\rangle -C| s\rangle ),
\end{eqnarray}
respectively. Capital letter $S=\sqrt{1-C^{2}}$, and $C =\langle r|s\rangle$, where the maximum value of the fringe visibility becomes the overlap between two linearly independent states.
In Eq.~(\ref{3eq-04}), parameters $\alpha$ and $\beta$ are chosen
to minimize the probability of failure
\begin{eqnarray}
Q &=& \omega _{b}Tr( \rho _{in}^{D}{\Pi }_{0}) +\omega
_{a}Tr( \rho_{out} ^{D}{\Pi }_{0}) .
\end{eqnarray}
By the Cauchy inequality and the resolution of the identity, we derive the lower bound on the probability of failure \cite{Feng}
\begin{eqnarray}
Q &\geq & 2\sqrt{ \omega _{a} \omega _{b}}F(\rho _{in}^{D},\rho_{out}^{D})  \label{3eq-05}, 
\end{eqnarray}
where the fidelity \cite{Nielsen} is defined as
\begin{equation}
F =Tr|\sqrt{\rho _{in}^{D}\rho_{out}^{D}}| = C.
\end{equation}
The lower bound of the failure probability is achieved if and only if
\begin{equation}
 \omega _{b}Tr( \rho _{in}^{D}\Pi _{0}) = \omega _{a}Tr( \rho_{out} ^{D}\Pi
_{0}) =\sqrt{ \omega _{a} \omega _{b}}F( \rho _{in}^{D},\rho_{out} ^{D}) \label{3eq-06}.
\end{equation}
\begin{figure*}[tbp]
\includegraphics[clip=true,height=8cm,width=14cm]{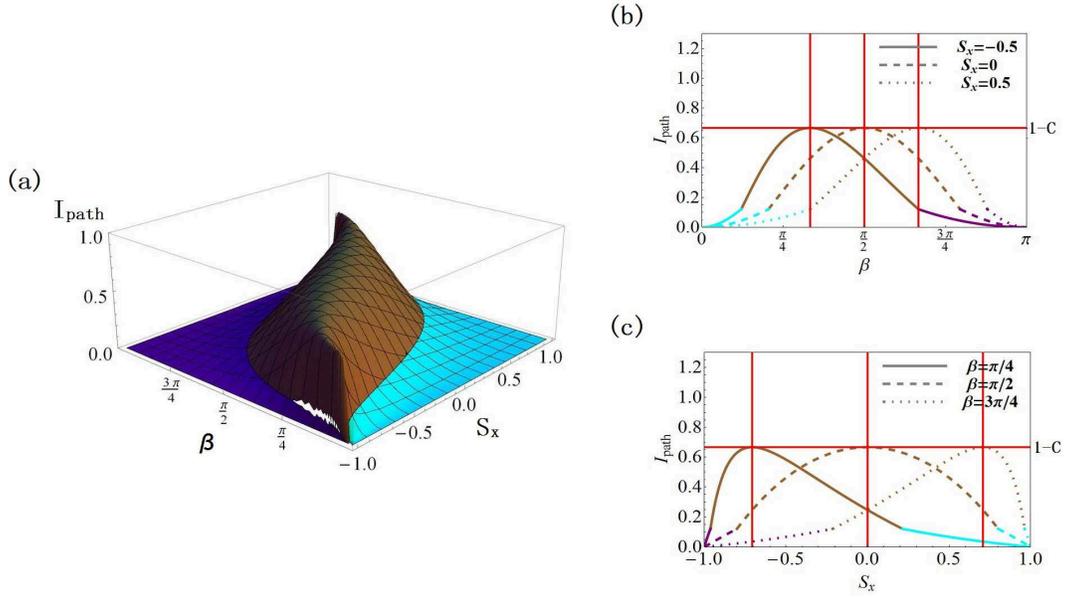}
\caption{(Color online). (a) The WPI as the function of $S_{x}$ and $\beta$ with $C=1/3$,
the range with purple, brown, and cyan represent the $I_{path}$ under the conditions $\sqrt{\omega _{a} / \omega _{b}} \in (0,C)$, $(C,C^{-1})$, and $(C^{-1},+\infty)$, respectively.
(b) The cross section of 3D surface for $S_{x}=-0.5, 0, 0.5$, (c) the cross section of 3D surface for $\beta=\pi/4, \pi/2, 3\pi/4$.} \label{fig4.eps}
\end{figure*}
\begin{figure*}[tbp]
\includegraphics[clip=true,height=8cm,width=14cm]{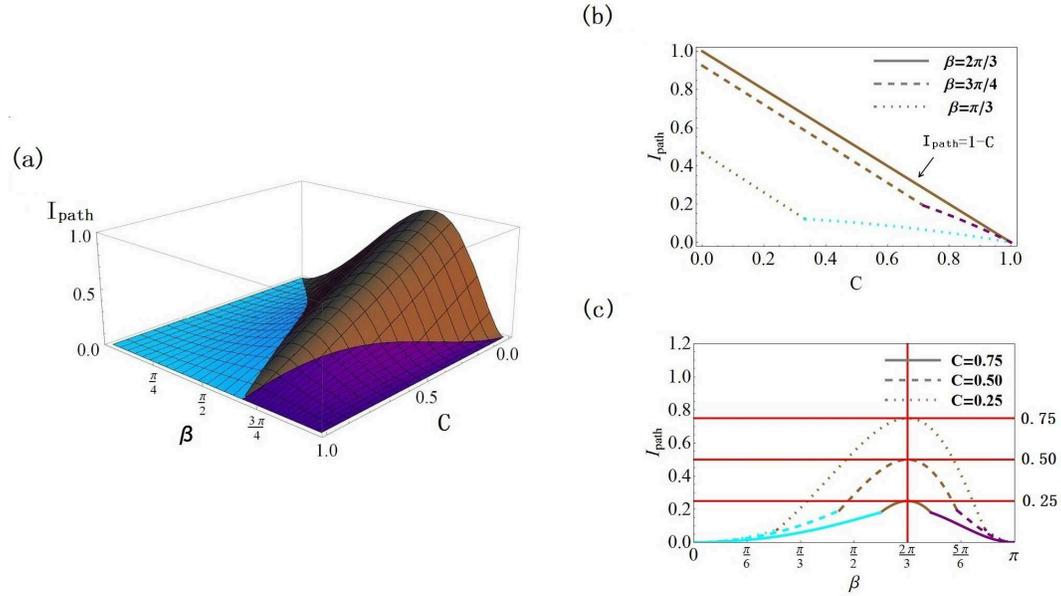}
\caption{(Color online). (a) The WPI as the function of $\beta$ and $C$ with $S_{x}=1/2$,
the range with purple, brown, and cyan represent the $I_{path}$ under the conditions $\sqrt{\omega _{a} / \omega _{b}} \in (0,C)$, $(C,C^{-1})$, and $(C^{-1},+\infty)$, respectively.
(b) The cross section of 3D surface for $\beta=2\pi/3, 3\pi/4, \pi/3$, (c) the cross section of 3D surface for $C=0.75, 0.50, 0.25$.} \label{fig5.eps}
\end{figure*}
\begin{figure*}[tbp]
\includegraphics[clip=true,height=8cm,width=14cm]{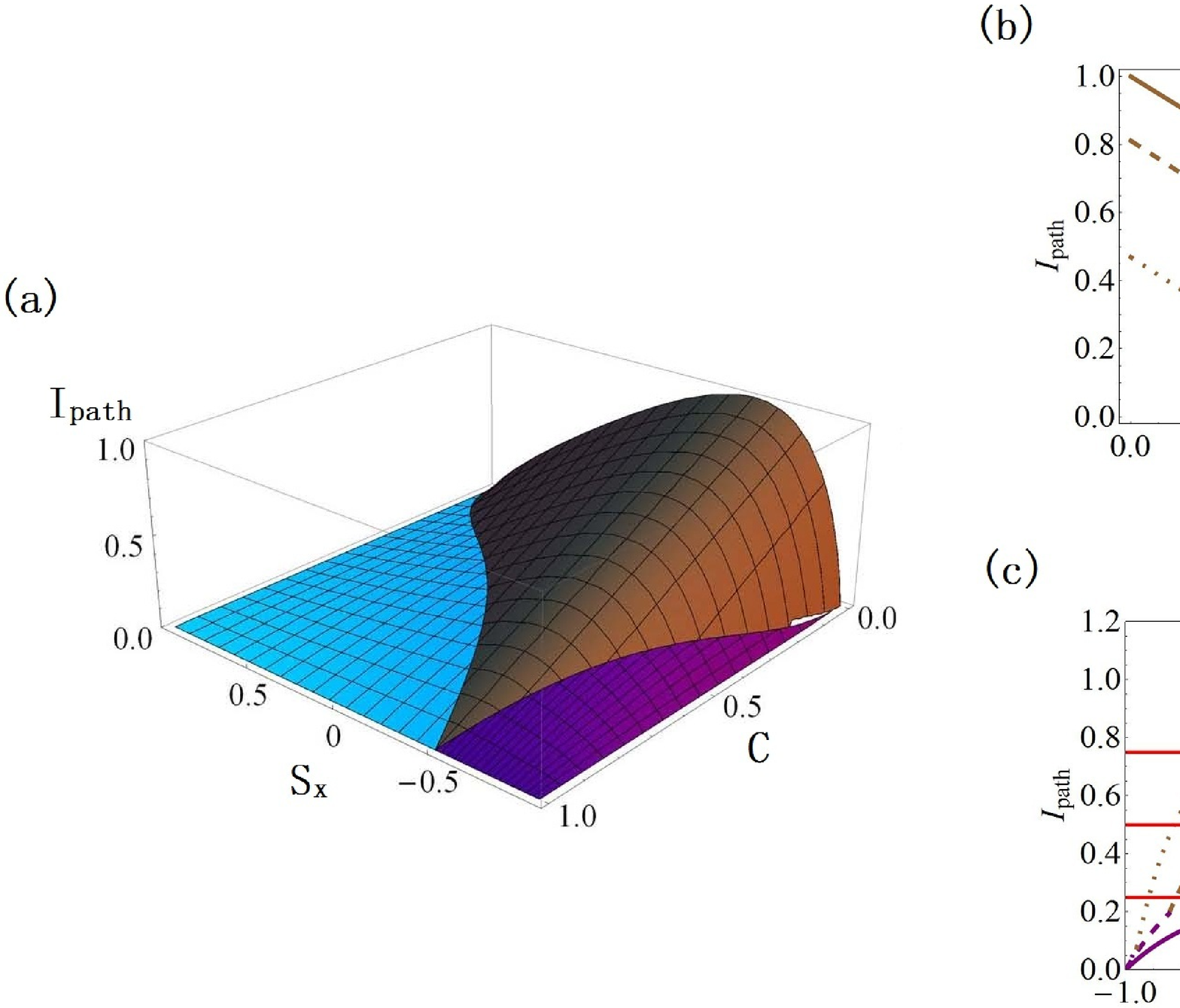}
\caption{(Color online). (a) The WPI as the function of $S_{x}$ and $C$ with $\beta=\pi/3$,
the range with purple, brown, and cyan represent the $I_{path}$ under the conditions $\sqrt{\omega _{a} / \omega _{b}} \in (0,C)$, $(C,C^{-1})$, and $(C^{-1},+\infty)$, respectively.
(b) The cross section of 3D surface for $S_{x}=-0.5, 0, 0.5$, (c) the cross section of 3D surface for $C=0.75, 0.50, 0.25$.} \label{fig5.eps}
\end{figure*}

The probability of the kth outcome, $tr(\Pi _{k}\rho _{k}^{D})$, is always real and nonnegative, which requires $0\leq\alpha,\beta\leq1$. The choice of the measurement that discriminates
$| r\rangle$ and $| s\rangle$ unambiguously depends on the relation between the ratio $\sqrt{\omega _{a} / \omega _{b}}$ and the overlap C:

(1) When $\sqrt{\omega _{a} / \omega _{b}} \leq C$, the minimum probability of failure $Q=\omega _{a}+C^{2}\omega _{b}$ is achieved by
selecting the following measurement operators $\Pi _{a} =0$, $\Pi _{b} =| s^{\perp}\rangle\langle s^{\perp}|$, $\Pi _{0} =| s\rangle\langle s|$.
Here, state $| s\rangle$ is never detected, and the optimal POVM becomes a von Neumann projective measurement. Actually, in this case, $\omega _{b}>\omega _{a}$, the state of the detector is more likely to be
in state $\rho _{in}^{D}$, so we make the failure direction along $\rho _{out}^{D}$ for obtaining the minimum probability of failure.
The joint probability is obtained as                                                                                                                                                                                                 \begin{eqnarray}
Q( b,b)&=&(1-C ^{2})\frac{\sin ^{2}\frac{\beta }{2}( 1-S_{x}) }{1+S_{x}\cos \beta },
\end{eqnarray}
\begin{eqnarray}
Q( a,b)&=& 0,
\end{eqnarray}
\begin{eqnarray}
Q( a,0)&=&\frac{\cos^{2}\frac{\beta }{2}( 1+S_{x}) }{1+S_{x}\cos \beta },
\end{eqnarray}
\begin{eqnarray}
Q( b,0)&=& C ^{2}\frac{\sin ^{2}\frac{\beta }{2}( 1-S_{x}) }{1+S_{x}\cos \beta }.
\end{eqnarray}
Then, the amount of WPI via the von Neumann projective measurement reads
\begin{eqnarray}
I_{path}&=& ( 1- C^{2})\frac{\sin ^{2}\frac{\beta }{2}( 1-S_{x}) }{1+S_{x}\cos \beta }
\log \frac{1+S_{x}\cos \beta }{\sin ^{2}\frac{\beta }{2}(
1-S_{x}) } \label{3eq-08}.
\end{eqnarray}

(2) When $C\leq \sqrt{\omega _{a} / \omega _{b}}\leq 1/C$, the minimum probability of failure $Q=2C\sqrt{ \omega _{a} \omega _{b}}$
is achieved by choosing the following measurement operators
\begin{eqnarray}
\Pi_{a} &=&\frac{1}{S^{2}}( 1-C\tan
\frac{\beta }{2}\sqrt{\frac{1-S_{x}}{1+S_{x}}}) | r^{\perp}\rangle \langle r^{\perp}|  ,
\end{eqnarray}
\begin{eqnarray}
\Pi_{b} &=&\frac{1}{S^{2}}( 1-C\cot
\frac{\beta }{2}\sqrt{\frac{1+S_{x}}{1-S_{x}}}) |
s^{\perp}\rangle \langle s^{\perp}| ,
\end{eqnarray}
\begin{eqnarray}
\Pi_{0} &=& C\cot \frac{\beta }{2}\sqrt{%
\frac{1+S_{x}}{1-S_{x}}}| r\rangle \langle
r|  \notag \\
&&+\frac{ C}{S}( 1-C\cot \frac{\beta }{2}\sqrt{\frac{%
1+S_{x}}{1-S_{x}}})| r\rangle
\langle r^{\perp}|   \notag \\ &&+\frac{C}{S}( 1-C\cot \frac{\beta }{2}\sqrt{\frac{1+S_{x}}{1-S_{x}}%
}) | r^{\perp}\rangle \langle
r|   \notag \\
&&+[1-\frac{C^{2}}{S^{2}}( 1-C\cot \frac{\beta }{2}\sqrt{\frac{%
1+S_{x}}{1-S_{x}}}) \notag \\
&&-\frac{1}{S^{2}}( 1-C\tan \frac{\beta
}{2}\sqrt{\frac{1-S_{x}}{1+S_{x}}}) ]| r^{\perp}\rangle \langle r^{\perp}|  ,
\end{eqnarray}
This measurement is more general than the von Neumann projective measurement. Via Eq.~(\ref{3eq-02}), the joint probability reads
\begin{eqnarray}
Q( a,a)&=&\frac{\cos ^{2}\frac{\beta }{2}( 1+S_{x}) }{1+S_{x}\cos \beta }%
( 1-C\tan \frac{\beta }{2}\sqrt{\frac{1-S_{x}%
}{1+S_{x}}}),
\end{eqnarray}
\begin{eqnarray}
Q( b,b)&=&\frac{\sin ^{2}\frac{\beta }{2}( 1-S_{x}) }{1+S_{x}\cos \beta }%
( 1-C\cot \frac{\beta }{2}\sqrt{\frac{1+S_{x}%
}{1-S_{x}}}),
\end{eqnarray}
\begin{equation}
Q( a,b)=Q( b,a)=0,
\end{equation}
\begin{eqnarray}
Q( a,0)&=& C\frac{\cos ^{2}\frac{\beta }{2}( 1+S_{x}) }{1+S_{x}\cos \beta }%
 \tan \frac{\beta }{2}\sqrt{\frac{1-S_{x}}{1+S_{x}}},
\end{eqnarray}
\begin{eqnarray}
Q( b;0)&=& C\frac{\sin ^{2}\frac{\beta }{2}( 1-S_{x}) }{1+S_{x}\cos \beta }%
\cot \frac{\beta }{2}\sqrt{\frac{1+S_{x}}{
1-S_{x}}}.
\end{eqnarray}
Then the amount of WPI obtained from the POVM measurement is calculated as
\begin{eqnarray}
I_{path}&=&\frac{\cos ^{2}\frac{\beta }{2}( 1+S_{x}) }{1+S_{x}\cos \beta }%
( 1-C\tan \frac{\beta }{2}\sqrt{\frac{1-S_{x}}{1+S_{x}}}) \notag\\
&&\times\log \frac{1+S_{x}\cos \beta }{\cos ^{2}\frac{\beta }{2}( 1+S_{x}) }  \notag\\
&&+\frac{\sin ^{2}\frac{\beta }{2}( 1-S_{x}) }{1+S_{x}\cos \beta }%
( 1-C\cot \frac{\beta }{2}\sqrt{\frac{1+S_{x}}{1-S_{x}}}) \notag\\
&&\times\log \frac{1+S_{x}\cos \beta }{\sin ^{2}\frac{\beta }{2}( 1-S_{x}) } \label{3eq-09},
\end{eqnarray}
according to its definition given by Eq.~(\ref{3eq-03}).

(3) When $\sqrt{\omega _{a} / \omega _{b}} \geq 1/C$, the minimum probability of failure $Q=\omega _{b}+C^{2}\omega _{a}$ is achieved by
selecting the measurement operators $\Pi _{a} =| r^{\perp}\rangle\langle r^{\perp}|$, $\Pi _{b} =0$, $\Pi _{0} =| r\rangle \langle r|$.
Here, state $| r\rangle$ is never detected, and the optimal POVM becomes a von Neumann projective measurement. In this case, $\omega _{a}>\omega _{b}$. The state of the detector is more likely to be
in state $\rho _{out}^{D}$, so the failure direction is chosen along $\rho _{in}^{D}$ for obtaining the minimum probability of failure.
The joint probability is obtained as
\begin{eqnarray}
Q( a,a)&=& (1-C ^{2})\frac{\cos ^{2}\frac{\beta }{2}( 1+S_{x}) }{1+S_{x}\cos \beta } ,
\end{eqnarray} \begin{equation}
Q( b,a)= 0,
\end{equation}
\begin{eqnarray}
Q( a,0)&=& C ^{2}\frac{\cos ^{2}\frac{\beta }{2}( 1+S_{x}) }{1+S_{x}\cos \beta },
\end{eqnarray}
\begin{eqnarray}
Q( b,0)&=&\frac{\sin ^{2}\frac{\beta }{2}( 1-S_{x}) }{1+S_{x}\cos \beta }.
\end{eqnarray}
And, the amoumt of WPI
is given by
\begin{eqnarray}
I_{path}&=& ( 1-C ^{2}) \frac{\cos ^{2}\frac{\beta }{2}( 1+S_{x}) }{1+S_{x}\cos \beta } \log \frac{1+S_{x}\cos \beta }{\cos ^{2}\frac{\beta }{2}(
1+S_{x}) } \label{3eq-10}.
\end{eqnarray}

Eqs.~(\ref{3eq-08}), (\ref{3eq-09}) and (\ref{3eq-10}) show that the WPI is a piecewise function of the parameters $S_{x}$, $\beta$, and C.
Although all the components of the Bloch vector determines the fringe visibility, only $S_{x}$ occurs in the expression of the WPI, indicating that
$I_{path}$ is independent of the initial state of the quantum particle. In Fig.4, we plot the WPI as the function of $S_{x}$ and $\beta$ with
the overlap $C=1/3$. The $I_{path}$ under the conditions $\sqrt{\omega _{a} / \omega _{b}} \in (0,C)$, $(C,C^{-1})$, and $(C^{-1},+\infty)$ is shown in the ranges
with purple, brown, and cyan in Fig.4(a), respectively. The white range in Fig.4(a) indicates that $I_{path}$ is a discontinuous function of $S_{x}$ and $\beta$.
It can be observed from Fig.4 that $I_{path}\leq1-C$ for any $S_{x}$ and $\beta$, and the position along the $S_{x}(\beta)$ axis that $I_{path}=1-C$ occurs varies with
different given $\beta(S_{x})$. In Fig.5(6), we have plotted $I_{path}$ as the function of parameters $\beta(S_{x})$ and C for a given $S_{x}(\beta)$.
One can also find that $I_{path}$ is less than or equal to $1-C$, $I_{path}$ decrease as C increase, the position that the peak occurs is fixed for different overlap C,
i.e., the peak appears at $\beta=2\pi/3$ when $S_{x}=1/2$ in Fig.5(c), and $S_{x}=-1/2$ when $\beta=\pi/3$ in Fig.6(c). From Eq.~(\ref{3eq-09}), we find that the
maximum of $I_{path}$ can be achieved once $\cos \beta=-S_{x}$.

The wave-like and the particle-like property are quantitated by fringe visibility V in Eq.~(\ref{2eq-02}) and the WPI $I_{path}$ in Eq.~(\ref{3eq-03}).
Since $V\leq C$ and $I_{path}\leq1-C$, we obtain the complementary relation
\begin{eqnarray}
V+I_{path}\leq1   \label{3eq-11},
\end{eqnarray}
the equal sign holds in  Eq.~(\ref{3eq-11}) when $\cos \beta=-S_{x}$.

\section{\label{Sec:4} conclusion}

We have investigated the complementarity of the fringe visibility and the WPI in a MZI with one asymmetric BS. Although the fringe visibility measured in either two output ports are
different, there exists an upper limit, i.e. $V\leq C$. The upper bound $C=|tr_{D}(U\rho_{in} ^{D})|$ is determined by the initial state of the detector and the unitary operator performed on the detector. The
maximum value of the fringe visibility C can be achieved when the quantum system is initially in pure state with $\cos \beta=-S_{x}$. To observe the particle-like behavior of this quantum system, a four-path
interferometer must be introduced due to the asymmetrical BS2. The WPI is characterized by the WPI $I_{path}$, which is obtained via the unambiguous discrimination on the
state of the detector. Although $I_{path}$ is dependent on the asymmetric BS and the initial state of the quantum system, the WPI is bounded by the following inequality, $I_{path}\leq1-C$.
The maximum $I_{path}$ is achieved when $\cos \beta=-S_{x}$. It is also found that $V+I_{path}\leq1$.

\begin{acknowledgments}

This work was supported by NSFC Grants No. 11374095,
No. 11422540, No. 11434011, No. 11575058; National
Fundamental Research Program of China (the 973 Program)
Grant No. 2012CB922103; Hunan Provincial Natural Science
Foundation of China Grants No. 11JJ7001.
\end{acknowledgments}

\end{document}